\documentclass[8.5pt,twoside,twocolumn]{article} 
\usepackage[super,sort&compress,comma]{natbib}  
\usepackage{mhchem} 
\usepackage{times,mathptmx,bm,multirow} 
\usepackage{sectsty} 
\usepackage{balance}  
 
\usepackage{graphicx} 
\usepackage{epstopdf} 
\usepackage{amsmath} 
\usepackage{lastpage} 
\usepackage[format=plain,justification=raggedright,singlelinecheck=false,font=small,labelfont=bf,labelsep=space]{caption}  
\usepackage{fancyhdr} 
\begin{document} 
 
\def\openone{\leavevmode\hbox{\small1\kern-3.3pt\normalsize1}} 
\def\half{ \frac{1}{2}} 
\def\eps{\varepsilon} 
\def\vphi{\varphi} 

\newcommand{\Op}[1]{\mathbf{\hat{#1}}} 
\newcommand{\vOp}[1]{\mathbf{\hat{\vec{#1}}}} 
\newcommand{\dotOp}[1]{\dot{\hat{\mathsf{\boldsymbol{#1}}}}} 
\newcommand{\ddotOp}[1]{\ddot{\hat{\mathsf{\boldsymbol{#1}}}}} 
\newcommand{\Opnohat}[1]{\boldsymbol{\mathsf{#1}}} 
\newcommand{\Fkt}[1]{\,\mathsf {#1}} 
\newcommand{\e}{\Fkt{e}} 
\ifx\Tr\renewcommand{\Tr}{\Fkt{Tr}} 
\else\newcommand{\Tr}{\Fkt{Tr}} 
\fi 
\newcommand{\diff}{\Fkt{d}} 
\newcommand{\sub}[1]{_{\mathsf {#1}}} 
\newcommand{\up}[1]{^{\mathsf {#1}}} 
\renewcommand{\dagger}{+} 
\newcommand\hateq{\ensuremath{\hat =}} 
\newcommand{\clc}[1]{\multicolumn{1}{c}{#1}} 
\newcommand{\be}{\begin{equation}} 
\newcommand{\ee}{\end{equation}} 
\newcommand{\rff}[1]{(\ref{#1})} 

\thispagestyle{plain} 
\fancypagestyle{plain}{ 
\renewcommand{\headrulewidth}{1pt}} 
\renewcommand{\thefootnote}{\fnsymbol{footnote}} 
\renewcommand\footnoterule{\vspace*{1pt}%
\hrule width 3.4in height 0.4pt \vspace*{5pt}}  
\setcounter{secnumdepth}{5} 

\makeatletter
\DeclareRobustCommand\onlinecite{\@onlinecite}
\def\@onlinecite#1{\begingroup\let\@cite\NAT@citenum\citealp{#1}\endgroup}
\makeatother   

\makeatletter  
\def\subsubsection{\@startsection{subsubsection}{3}{10pt}{-1.25ex plus -1ex minus -.1ex}{0ex plus 0ex}{\normalsize\bf}}  
\def\paragraph{\@startsection{paragraph}{4}{10pt}{-1.25ex plus -1ex minus -.1ex}{0ex plus 0ex}{\normalsize\textit}}  
\renewcommand\@biblabel[1]{#1}             
\renewcommand\@makefntext[1]%
{\noindent\makebox[0pt][r]{\@thefnmark\,}#1} 
\makeatother  
\renewcommand{\figurename}{\small{Fig.}~} 
\sectionfont{\large} 
\subsectionfont{\normalsize}  
 
\fancyfoot{} 
\fancyfoot[RO]{\footnotesize{\sffamily{1--\pageref{LastPage} ~\textbar  \hspace{2pt}\thepage}}} 
\fancyfoot[LE]{\footnotesize{\sffamily{\thepage~\textbar\hspace{3.45cm} 1--\pageref{LastPage}}}} 
\fancyhead{} 
\renewcommand{\headrulewidth}{1pt}  
\renewcommand{\footrulewidth}{1pt} 
\setlength{\arrayrulewidth}{1pt} 
\setlength{\columnsep}{6.5mm} 
\setlength\bibsep{1pt} 
\twocolumn[ 
  \begin{@twocolumnfalse} 
\noindent\LARGE{\textbf{Formation of ultracold SrYb molecules in an 
    optical lattice by photoassociation spectroscopy: theoretical 
    prospects}}  
\vspace{0.6cm} 
 
\noindent\large{\textbf{Micha\l~Tomza,\textit{$^{a}$} Filip 
    Paw\l{}owski,\textit{$^{a,b}$} Ma\l{}gorzata 
    Jeziorska,\textit{$^{a}$}   Christiane P. Koch,\textit{$^{c}$}  
    and Robert Moszynski\textit{$^{a}$}}}\vspace{0.5cm}  
 
 
\vspace{0.6cm} 
 
\noindent \normalsize{ 
State-of-the-art {\em ab initio} techniques have been applied to compute the
potential energy curves for the SrYb molecule in the Born-Oppenheimer
approximation
for the ground state and first fifteen excited singlet and triplet states.
All the excited state potential energy curves were computed using
the equation of motion approach
within the coupled-cluster singles and doubles framework and large basis-sets,
while the ground state potential was computed using the coupled cluster
method with single, double, and noniterative triple excitations. The leading
long-range coefficients describing the dispersion interactions
at large interatomic distances are also reported.
The electric transition dipole moments
have been obtained as the first residue of the
polarization propagator computed with the linear response coupled-cluster
method restricted to single and double excitations.
Spin-orbit coupling matrix elements have been evaluated using the
multireference configuration interaction method restricted to single
and double excitations with a large active space.
The electronic structure 
data was employed to investigate the possibility of forming  
deeply bound ultracold SrYb molecules in an optical lattice in a 
photoassociation experiment using continuous-wave lasers.  
Photoassociation near the intercombination line transition of
atomic strontium into the vibrational levels of the 
strongly spin-orbit mixed $b^3\Sigma^+$, $a^3\Pi$, $A^1\Pi$, 
and $C^1\Pi$ states with subsequent efficient stabilization  
into the $v^{\prime\prime}=1$ vibrational level of the electronic ground state is 
proposed. Ground state SrYb molecules can be accumulated by making use of 
collisional decay from $v^{\prime\prime}=1$ to $v^{\prime\prime}=0$. 
Alternatively, photoassociation and stabilization to $v^{\prime\prime}=0$ 
can proceed via stimulated Raman adiabatic passage provided that the 
trapping frequency of the optical lattice  
is large enough and phase coherence between the 
pulses can be maintained over at least tens of microseconds.  
} 
\vspace{0.5cm} 
 \end{@twocolumnfalse} 
  ] 
 
\footnotetext{\textit{$^{a}$~Faculty of Chemistry, University of Warsaw, Pasteura 1, 02-093 Warsaw, Poland.}} 
\footnotetext{\textit{$^{b}$~Physics Institute, Kazimierz Wielki University, pl. Weyssenhoffa 11, 85-072 Bydgoszcz, Poland. }} 
\footnotetext{\textit{$^{c}$~Theoretische Physik, Universit\"at Kassel, Heinrich-Plett-Str. 40, 34132 Kassel, Germany.}} 
 
 
\section{\label{sec:0}Introduction} 
 
Molecules cooled to temperatures below $T=10^{-3}\,$K allow for 
tackling questions touching upon the very fundamentals of quantum 
mechanics. They are also promising candidates in novel applications, 
ranging from ultracold chemistry and precision measurements to quantum 
computing. Cold and ultracold molecules are thus  
opening up new and exciting areas of research in chemistry and 
physics. Due to  
their permanent electric dipole moment, polar molecules are  
particularly interesting objects of study:  
Dipole-dipole interactions are long range and  
can precisely be controlled with external electric fields 
turning the experimental parameters field strength and orientation 
into the knobs that control the quantum dynamics of these molecules.  
It is not surprising then that a 
major objective for present day experiments on cold molecules 
is to achieve quantum degeneracy for polar molecules. 
Two approaches to this problem are used -- indirect methods, in which 
molecules are formed from pre-cooled atomic gases,\cite{Julienne:87,JonesRMP06,KohlerRMP06,NiScience08} 
and direct methods, in which molecules are cooled from molecular
beam temperatures, typically tens of Kelvin.\cite{Doyle:98,Meijer:99,Bethlem:03,Schnell:09,Meerakker:09} 
 
Direct cooling techniques, based on buffer gas cooling\cite{Doyle:98} 
or Stark deceleration \cite{Meijer:99}, produce cold molecules with a 
temperature well below 1~K. However,  a second-stage cooling 
process is required to reach  
temperatures below $10^{-3}\,$K. 
The most promising second-stage technique is  
sympathetic cooling where  cold molecules are introduced into 
an ultracold atomic gas and equilibrate with it. Sympathetic 
cooling has successfully  been used to achieve Fermi 
degeneracy in $^6$Li\cite{DeMarco:1999} and Bose-Einstein 
condensation in $^{41}$K\cite{Modugno:2001} and to obtain ultracold 
ions. \cite{Zipkes2010,Zipkes2010a,Schmid,KrychPRA11} 
For molecular systems, however, sympathetic cooling  
has not yet been attempted, and there are 
certainly many challenges to overcome. In fact, calculations 
of the scattering cross sections for the collisions of molecules 
that can be decelerated with 
ultracold coolant atoms suggest that   
sympathetic cooling will not be efficient in many 
cases.\cite{Lara:07,ZuchowskiPRA09,Tokunaga:11} 
 
Alternatively, direct methods first cool atoms to ultralow temperatures and then 
employ photoassociation\cite{JonesRMP06} or 
magnetoassociation\cite{KohlerRMP06}  
to create molecules, reaching translational 
temperatures of the order of a few $\mu$K or nK.  
In particularly fortuitious cases, photoassociation may directly produce 
molecules in their vibrational ground state.\cite{DeiglmayrPRL08},  
Typically, however, the molecules are created in extremely weakly bound 
levels, and follow-up stabilization to the ground state is necessary.  
For molecules built of alkali metal atoms, 
this has  been achieved using stimulated emission 
pumping~\cite{SagePRL05} or alternatively, 
 employing coherent control techniques such as  
Stimulated Raman Adiabatic Passage  
(STIRAP)~\cite{LangPRL08,NiScience08,OspelkausNatPhys08,DanzlSci08}     
and vibrational cooling of molecules with amplitude-shaped broadband 
laser light.~\cite{PilletSci08}  
 
Closed-shell atoms such as alkali earth metals are more challenging to 
cool and trap than open-shell atoms such as the alkalis. 
Closed-shell atoms do not have a magnetic moment in their ground state 
that enables magnetic trapping. Moreover, for alkaline earth metals  
the short lifetime of the first excited $^1P$ state implies rather 
high Doppler temperatures, making dual-stage cooling a necessity where 
the second stage operates near an intercombination line.  
Despite these obstacles, cooling of calcium, strontium, and ytterbium atoms to 
micro-Kelvin temperatures has been achieved, and Bose-Einstein 
condensates of $^{40}$Ca,\cite{KraftPRL09} 
$^{84}$Sr,\cite{StellmerPRL09,EscobarPRL09}  
$^{86}$Sr,\cite{StellmerPRA10}, $^{88}$Sr,\cite{MickelsonPRA10}  
$^{170}$Yb,\cite{FukuharaPRA07} and $^{174}$Yb \cite{TakasuPRL03} 
have been obtained. 
 
In contrast to alkali metal dimers\cite{KohlerRMP06} or molecules 
consisting of an alkali metal atom 
plus a closed-shell atom,\cite{ZuchowskiPRL10}  
the magnetoassociation of two closed-shell atoms is not feasible  
experimentally even if the nuclear spin is non-zero. 
The zero-field splittings and couplings between 
the atomic threshold and molecular states provided by the largest non-zero 
terms in the fine structure and hyperfine structure Hamiltonian for the 
electronic ground state, i.e., the scalar and tensor interactions 
between the nuclear magnetic dipole moments\cite{Brown03},  
are simply too small.\cite{KohlerRMP06}  
On the other hand, the closed-shell structure of the alkali earth 
metal and ytterbium atoms leads to very simple  molecular potentials 
with low radiative losses and weak coupling to the environment.   
This opens new areas of possible applications, such as  
manipulation of the scattering properties with low-loss optical 
Feshbach resonances,\cite{CiuryloPRA05} 
high-resolution photoassociation spectroscopy  
at the intercombination line,\cite{TojoPRL06,ZelevinskyPRL06} 
precision measurements to test for a  time variation of the 
proton-to-electron mass ratio,\cite{ZelevinskyPRL08} 
quantum computation with trapped polar molecules,\cite{DeMillePRL02} and 
ultracold chemistry\cite{OspelkausScience10}. 
 
To the best of our knowledge, production of heteronuclear diatomic  
molecules built of closed-shell atoms has not yet been achieved 
experimentally. Also such processes have not yet been considered  
theoretically. Here we fill this gap and report a theoretical study of the 
photoassociative formation of heteronuclear diatomic molecules from 
two closed-shell atoms on the example of the SrYb molecule. Although 
the SrYb molecule may seem very exotic, especially for chemists, 
strontium and ytterbium atoms are promising candidates for producing  
molecules since they have both successfully been cooled and trapped.  
Moreover, both Sr and Yb have many stable isotopes. Such a diversity 
of stable isotopes is key to controlling the collisional 
properties of  bosonic molecules with no magnetic moments and 
hyperfine structure. For example,  
one can effectively tune the interatomic interactions by choosing the 
most suitable isotope   
to achieve scattering lengths appropriate for evaporative cooling. 
Last but not least, we consider photoassociative formation of SrYb 
molecules since there are 
ongoing experiments \cite{Zelevinsky} aiming at producing cold SrYb 
molecules in an optical lattice. 
 
The plan of our paper is as follows. Section~\ref{sec2} describes 
the theoretical methods used in the {\em ab initio} 
calculations and discusses the electronic structure of SrYb in terms 
of the ground and excited-state potentials,  
transition moments, spin-orbit and nonadiabatic angular 
couplings. Section~\ref{sec:3} analyzes the vibrational structure of 
the SrYb molecule as a prerequisite to determine an efficient route 
for photoassociation followed by stabilization into the vibronic 
ground state. It also discusses prospects of producing cold 
SrYb molecules by photoassociation near the intercombination line of 
strontium, and subsequent spontaneous or stimulated emission.  
Section \ref{sec:4} summarizes our findings.

\section{Electronic structure of SrYb} 
\label{sec2} 
In the present study we adopt the computational scheme successfully 
applied before to the ground and excited states of the calcium 
dimer\cite{BusseryPRA03,BusseryPRA05,BusseryJCP06,BusseryMP06,KochMoszPRA08}   
and of the (BaRb)$^+$ molecular ion.\cite{KrychPRA11} 
The potential energy curves for the ground and excited states of the SrYb molecule 
have been obtained by a supermolecule method:  
\be 
V^{^{2S+1}|\Lambda|}(R)= 
E_{\rm AB}^{\rm SM} - 
E_{\rm A}^{\rm SM}-E_{\rm B}^{\rm SM} 
\label{cccv} 
\ee 
where $E_{\rm AB}^{\rm SM}$ denotes 
the energy of the dimer computed using the supermolecule method SM, and $E_{\rm X}^{\rm SM}$,  
X=A or B, is the energy of the atom X. 
For the ground state potential we used the coupled cluster method restricted to single, double, and 
noniterative triple excitations, CCSD(T).  
Calculations on all the excited states 
employed the linear response theory (equation of motion) within the coupled-cluster singles 
and doubles (LRCCSD) framework~\cite{JorgensenJCP90}.  
The CCSD(T) and LRCCSD calculations were performed with the {\sc dalton} 
program~\cite{dalton}. Note that the methods used in our  
calculations are strictly size-consistent, so they ensure a proper dissociation 
of the electronic states, and a proper long-range asymptotics of the 
corresponding potential energy curves. 
This is especially important when dealing with collisions at ultralow temperatures, 
since the accuracy of the potential in the long range is crucial. 
 
For each electronic state we have also computed the long-range 
coefficients describing the dispersion interactions 
from the standard expressions (see, e.g. Refs.  
\onlinecite{Jeziorski94,Moszynski07}) 
that can be derived from the multipole expansion of the interatomic interaction operator. 
The long-range dispersion coefficients were computed with the 
recently introduced explicitly connected representation of the expectation value 
and polarization propagator within the coupled cluster method \cite{MoszynskiIJQC93,MoszynskiCzech05}, 
and the best approximation XCCSD4 proposed by Korona and collaborators \cite{KoronaMPhys06}. 
For the singlet and triplet states dissociating into Sr($^1P$)+Yb($^1S$), and 
Sr($^3P$)+Yb($^1S$), respectively, 
the dispersion coefficients were obtained from the sum-over-state 
expression with the transition moments and excitation energies computed with the 
multireference configuration interaction method limited to single and double 
excitations (MRCI).  
 
The transitions from the ground X$^1\Sigma^+$ state to the 
$^1\Sigma^+$ and  
$^1\Pi$ states are electric dipole allowed. 
The transition dipole moments for the electric, $d_i$,  
transitions were computed from the following expression 
\cite{Bunker98}: 
\be 
d_i=\langle X^1\Sigma^+|\Op{d}|(n)^{1}|\Lambda|\rangle, 
\label{trandipel} 
\ee 
where  
$\Op{d}$ denotes the dipole moment operator.
Note that in Eq. (\ref{trandipel}) $i=x$ or $y$ corresponds to transitions 
to $^1\Pi$ states, while $i=z$ corresponds to transitions to $^1\Sigma^+$ 
states.  
In the present calculations the electric transition dipole moments 
were computed as the first residue 
of the LRCCSD linear response function with two electric, $r$,  
operators~\cite{JorgensenJCP90}. In these calculations 
we have used the {\sc dalton} program~\cite{dalton}. 
We have evaluated the dependence of the transition dipole moments with 
the internuclear distance   
for the same set of distances as the excited state potential energy 
curves.  
 
\begin{figure}[tb] 
\centering 
\includegraphics[width=0.95\linewidth]{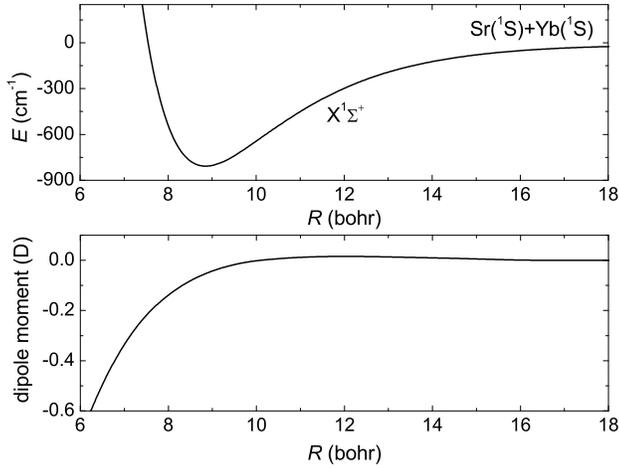} 
\caption{Potential energy curve (upper panel) and permanent dipole 
  moment (lower panel) of the $X^1\Sigma^+$ electronic ground state of 
  the SrYb molecule.}  
\label{fig:X1Sig} 
\end{figure} 
 
The electronic states of the 
low lying excited states of the SrYb molecule are coupled by nonadiabatic couplings. Therefore, in this work we have computed the most 
important angular coupling matrix elements: 
\be 
A(R)=\langle(n)^{2S+1}|\Lambda||L_+|(n')^{2S+1}|\Lambda'|\rangle. 
\label{ang} 
\ee 
In the above expression $L_+$ 
denotes the raising electronic angular momentum operator. Note that the electronic angular 
momentum operator couples states with $\Lambda$ differing by one. 
Nonadiabatic couplings were obtained with the MRCI method and 
the {\sc molpro} code \cite{molpro}. 
We have evaluated the dependence of the nonadiabatic coupling matrix elements with 
the internuclear distance 
for the same set of distances as the excited state potential energy 
curves. 

\begin{figure}[tb] 
\centering 
\includegraphics[width=0.95\linewidth]{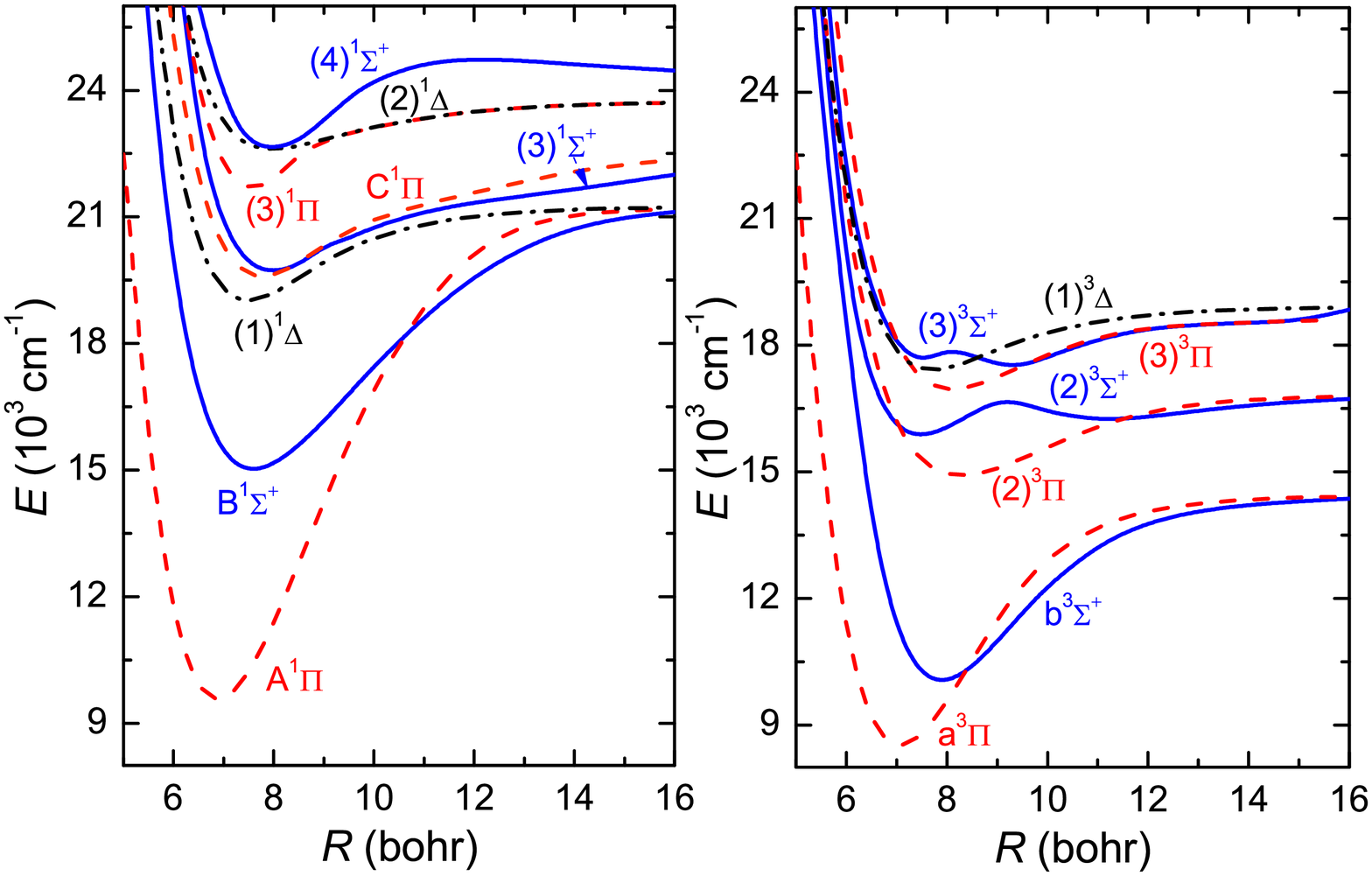} 
\caption{Potential energy curves of singlet (left) and triplet (right) excited states of SrYb dimer.} 
\label{fig:Allcurves} 
\end{figure} 

Strontium and ytterbium are heavy atoms, so  
the electronic states of the SrYb molecule are strongly mixed by the 
spin-orbit (SO) interactions. Therefore, in any analysis of the formation of 
the SrYb molecules 
the SO coupling and its dependence on the internuclear distance $R$ 
must be taken into account. 
We have evaluated the spin-orbit coupling matrix elements   
for the lowest dimer states 
that couple to the 0$^{+/-}$, 1, and 2 states of SrYb,  
with the spin-orbit coupling operator $H_{\rm SO}$ defined within the Breit-Pauli 
approximation \cite{Bethe57}.  
The spin-orbit coupling matrix elements  
have been  computed within the MRCI framework 
with the {\sc molpro} code~\cite{molpro}. 
Diagonalization of the relativistic Hamiltonian gives the spin-orbit coupled potential 
energy curves for the $0^{+/-}$, $1$ and $2$ states, respectively. Note that 
all potentials in the Hamiltonian matrices were 
taken from CCSD(T)  and LRCCSD calculations. Only the 
diagonal and nondiagonal spin-orbit coupling matrix elements were 
obtained with the MRCI method. 
Once the eigenvectors of these matrices are available, one can easily 
get the electric dipole transition moments and the nonadiabatic coupling 
matrix elements between the relativistic states. 
In order to mimic the scalar relativistic effects some electrons were described  
by pseudopotentials. For Yb we took the ECP28MWB pseudopotential \cite{KauppJCP91}, while  
for Sr the ECP28MDF \cite{LimJCP06} pseudopotential, both from the Stuttgart library.  
For the strontium and ytterbium atoms we used $spdfg$ quality basis sets \cite{LimJCP06,CaoJCP01},  
augmented with a set of $[2pdfg]$ diffuse functions.  
In addition, this basis set was augmented by the set of bond functions consisting of  
$[3s3p2d1f]$ functions placed in the middle of SrYb dimer bond.  
The full basis of the dimer was used in the supermolecule 
calculations and the Boys and Bernardi scheme was used 
to correct for the basis-set superposition error \cite{BoysMPhys70}. 
 
\begin{figure}[tb] 
\centering 
\includegraphics[width=0.95\linewidth]{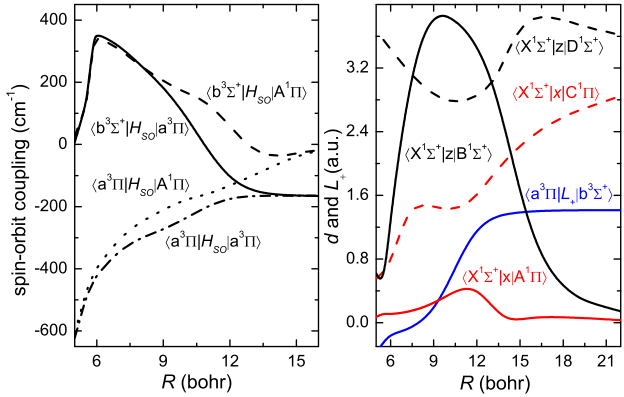} 
\caption{Left: matrix elements of the spin-orbit interaction for the $a^3\Pi$, 
  $b^3\Sigma^+$, and $A^1\Pi$ electronically excited states of SrYb. 
  Right: matrix elements of the 
  electric transition dipole moment from the  $X^1\Sigma^+$ ground 
  electronic state to $A^1\Pi$ state (solid curve) and to the $C^1\Pi$ 
  state (dot-dashed curve), and matrix elements of the  
  nonadiabatic angular coupling between the $a^3\Pi$ and 
  $b^3\Sigma^+$ states of SrYb (dashed curve).  
}  
\label{fig:SOcoup} 
\end{figure} 

Calculations were done for the ground state and first fifteen 
(eight singlet and seven triplet) excited states of SrYb. 
The singlet states correspond to the  
Sr($^1D$)+Yb($^1S$),  Sr($^1P$)+Yb($^1S$), Sr($^1S$)+Yb($4f^{13}5d6s^2$), and Sr($^1S$)+Yb($^1P$) 
dissociations, while triplet states to Sr($^3P$)+Yb($^1S$), Sr($^1S$)+Yb($^3P$),  Sr($^3D$)+Yb($^1S$). 
The  potential energies were calculated for twenty interatomic distances 
$R$ ranging from 5 to 50 bohr for each potential curve. The ground state potential
is presented on Fig. \ref{fig:X1Sig}, while 
the potential energy curves for the excited states  are plotted in Fig.~\ref{fig:Allcurves}. 
The spectroscopic characteristics of all these 
states are reported in Table \ref{tab:tab1}. 
The separated atoms energy for each state was set equal to the experimental  
value. Numerical values  of the potentials are available from the authors on request.  
 
Before going on with the discussion of the potentials let us note that the atomic 
excitation energies obtained from the CCSD calculations are accurate.  
Our predicted position of the nonrelativistic $^3P$ state of strontium is 14463 cm$^{-1}$, 
to be compared with the experimental value of 14705 cm$^{-1}$ \cite{nist} deduced from 
the positions of the states in the $P$ multiplet and the Land\'e rule.  
For the $^1D$ and $^1P$ states of Sr we obtain less than 5\% difference 
with the experimental values listed by NIST \cite{nist}.  
Also the life times of the $^3P_1$ and $^1P_1$ states of Sr are accurately reproduced. 
For the $^1P_1$ state we obtained 4.92 ns to be compared with the experimental value of 
5.22(3) ns \cite{NagelPRL05}. For the $^3P_1$ the theoretical and experimental numbers 
are 22 and 20 $\mu$s \cite{ZelevinskyPRL06}, respectively. Similarly good results were  
also obtained for the ytterbium 
atom. Such a good agreement between theory and experiment for the atoms gives us 
confidence that the molecular results will be of similar accuracy, i.e. a few percent 
off from the exact results. 
 
\begin{table}[tb]
\begin{center} 
\small 
  \caption{Spectroscopic characteristics (equilibrium distance, well depth, harmonic constant) of the non-relativistic electronic states  
  of $^{88}$Sr$^{174}$Yb dimer.} 
  \label{tab:tab1} 
  \begin{tabular*}{1\columnwidth}{lrrrrcl} 
    \hline 
    State            & $R_e$ (bohr) & $D_e$ (cm$^{-1}$) &  $\omega_e$(cm$^{-1}$) & Dissociation \\ 
    \hline 
    $X^1\Sigma^+$   &  8.78  &    828  &   32.8    &   Sr($^1S$)+Yb($^1S$)  \\ 
    $A^1\Pi$        &  6.84  &  11851  &   94.8    &   Sr($^1D$)+Yb($^1S$)  \\ 
    $B^1\Sigma^+$   &  7.54  &   5201  &   63.5    &   Sr($^1D$)+Yb($^1S$)  \\ 
    $(1)^1\Delta$   &  7.42  &   1202  &   62.5    &   Sr($^1D$)+Yb($^1S$)  \\ 
    $(3)^1\Sigma^+$ &  7.91  &   2963  &   48.5    &   Sr($^1P$)+Yb($^1S$)  \\ 
    $(2)^1\Pi$      &  7.70  &   3112  &   61.6    &   Sr($^1P$)+Yb($^1S$)  \\    
    $(4)^1\Sigma^+$ &  7.84  &   1790  &   58.6    &   Sr($^1P$)+Yb($\frac{7}{2},\frac{3}{2}$) \\ 
    $(3)^1\Pi$      &  7.53  &   2153  &   72.5    &   Sr($^1P$)+Yb($\frac{7}{2},\frac{3}{2}$) \\    
    $(2)^1\Delta$   &  7.89  &   1175  &   40.2    &   Sr($^1P$)+Yb($\frac{7}{2},\frac{3}{2}$) \\ 
    a$^3\Pi$        &  7.02  &   6078  &   84.7    &   Sr($^3P$)+Yb($^1S$)  \\ 
    b$^3\Sigma^+$   &  7.84  &   4493  &   71.3    &   Sr($^3P$)+Yb($^1S$)  \\ 
    $(2)^3\Sigma^+$ &  7.39  &   1024  &   61.7    &   \multirow{2}*{Sr($^1S$)+Yb($^3P$)}  \\ 
$2^{\mathrm{nd}}$ min.& 11.02 &   622  &   21.0    &                        \\ 
    $(2)^3\Pi$      &  8.23  &   1947  &   42.4    &   Sr($^1S$)+Yb($^3P$)  \\ 
    $(3)^3\Sigma^+$ &  7.45  &    982  &   92.7    &   \multirow{2}*{Sr($^3D$)+Yb($^1S$)}  \\ 
$2^{\mathrm{nd}}$ min.& 9.33 &   1077  &   47.8    &                        \\ 
    $(3)^3\Pi$      &  8.04  &   1678  &   47.9    &   Sr($^3D$)+Yb($^1S$)  \\ 
    $(1)^3\Delta$   &  7.65  &   1422  &   50.8    &   Sr($^3D$)+Yb($^1S$)  \\ 
    \hline 
  \end{tabular*} 
\end{center} 
\end{table} 
\begin{table}[tb] 
\caption{Long-range dispersion coefficients (in a.u.) 
for ground and relevant excited states of the SrYb dimer.\label{tab2}} 
\begin{tabular}{lcc} 
\hline 
State  &$C_6$ & $C_8$  \\ 
\hline 
$X^1\Sigma^+$  &  2 688   &  294 748    \\ 
$A^1\Pi$       &  3 771   &  502 070    \\ 
$a^3\Pi$       &  1 265   &  509 068    \\ 
$b^3\Sigma^+$  &  6 754   &  317 656    \\ 
\hline 
\end{tabular} 
\end{table} 
 
The ground $X^1\Sigma^+$ state potential energy curve is presented in Fig. \ref{fig:X1Sig}.  
It follows from the naive molecular orbital theory that the SrYb molecule in the ground state  
should be considered as a van der Waals molecule since the molecular configuration has an equal  
number of bonding and antibonding electrons. No regular chemical bond is 
expected, except for a weak dispersion attraction and exchange-repulsion. 
Indeed, the ground state potential is weakly bound with the binding energy of 828 cm$^{-1}$.  
For $J=0$ it supports $N_\nu=62$ vibrational levels for the lightest isotope pair, up to  
$N_\nu=64$ for the heaviest isotopes. The changes of the number of bound rovibrational levels  
and of the position of the last vibrational level for different isotopes results in  
the changes in the scattering length from very negative to very positive values.  
This should allow to choose isotopes most suitable for cooling and manipulation.  
The equilibrium distance, well depth, harmonic frequency, and the long-range coefficients  
of the $X^1\Sigma^+$ state are reported in Table \ref{tab:tab1}. The permanent  
dipole moment of SrYb in ground electronic state as a function of the interatomic 
distance $R$ is presented in Fig. \ref{fig:X1Sig}. Except for short interatomic distances, the 
dipole moment is very small. This is not very surprising since the two atoms  
have very similar electronegativities and the charge flow from one atom to the 
other, after the formation of the weak van der Waals bond, is very small. In fact, 
similarly as the bonding of the ground state, the dipole moment of SrYb should be 
considered as a dispersion dipole \cite{ByersMP73}. At large interatomic distances it 
vanishes as $R^{-6}$ \cite{ByersMP74,MoszynskiMP96}. 
The vibrationally averaged dipole moment of SrYb in ground vibrational state 
is indeed very small and equal to 0.023 a.u. 
 
Potential energy curves of the excited singlet and triplet states of SrYb are 
presented in Fig. \ref{fig:Allcurves}. 
An inspection of Fig. \ref{fig:Allcurves} shows that the 
potential energy curves for the excited states of the SrYb molecule  are 
smooth with well defined minima. The potential energy curves of the (2) and 
(3)$^3\Sigma^+$ states show an avoided crossing,  
and exhibit a double minimum structure. 
These double minima on the potential energy curves 
are due to the strong nonadiabatic interactions between these states.  
Other potential energy curves do not show any unusual features, except for the 
broad maximum on the potential of the (4)$^1\Sigma^+$ which is most likely due to 
the interaction with a still more excited state not reported in the present work. 
Except for the shallow double minima of the (2)$^3\Sigma^+$ and (3)$^3\Sigma^+$ states, 
and shallow $\Delta$ states, 
all other excited states of the SrYb molecule are strongly bound with binding 
energies $D_e$ ranging from 
approximately 1790 cm$^{-1}$ for the (4)$^1\Sigma^+$ state up to as much as 
11851 cm$^{-1}$ for the A$^1\Pi$ state.  
 
Let us compare the pattern of the potential energy curves of the heteronuclear SrYb molecule 
with the homonuclear Sr$_2$ dimer \cite{CzuchajCPL03}. 
In general, molecular orbitals constructed from  
the linear combinations of the 
Sr($5p$)+Yb($6p$) or  Sr($4d$)+Yb($5d$) 
atomic orbitals are expected to have less bonding or antibonding 
character than the molecular orbitals constructed from the Sr($5p$)+Sr($5p$) 
or Sr($4d$)+Sr($4d$) atomic orbitals, because large atomic orbital energy 
differences make combination of these orbitals less effective. 
This would explain why many potential energy curves of the SrYb dimer are 
less attractive than the corresponding potential energy curves 
of the Sr$_2$ dimer. 
The strongly attractive character of the potential energy curves for the first 
$^3 \Sigma^+$ and the first  $^3 \Pi$ states converging in the long range to  
Sr($^3P$)+Yb($^1S$) 
asymptote could be a result of the stabilizing effect of the Yb($5d$) orbitals 
for the lowest unoccupied orbitals of the  $\sigma$ and $\pi$ 
symmetry (these molecular are combinations of the Yb($6p$) and Sr($5p$) orbitals), but also 
of the appropriate Yb($5d$) orbitals, closer in energy to Sr($5p$). 
Potential energy curves for 
the second $^1 \Sigma^+$ and second  $^1 \Pi$  states converging to 
Sr($^1P$)+Yb($^1S$) are less 
 attractive than the potential energy curves for 
the triplet states, similarly as for the corresponding states of the Sr$_2$ dimer. 
As for the Sr$_2$ dimer potential energy curves 
for the singlet $\Sigma^+$ and $\Pi$ states 
converging to the Sr($^1D$)+Yb($^1S$) asymptote 
have a much more attractive character than the triplet states 
converging to the Sr($^3D$)+Yb($^1S$) asymptote. 
 
\begin{figure}[tb] 
\centering 
\includegraphics[width=0.95\linewidth]{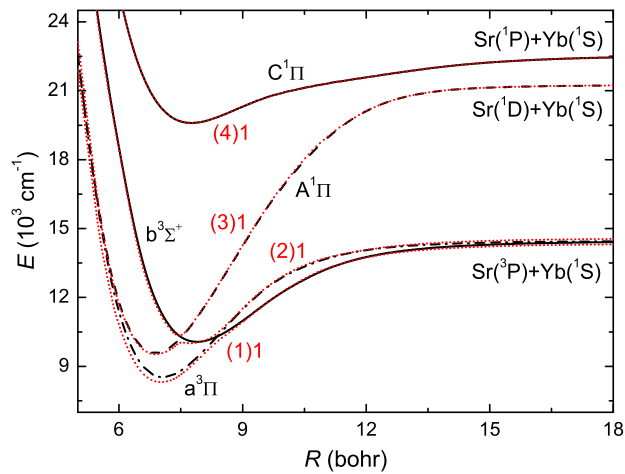} 
\caption{The $a^3\Pi$, $b^3\Sigma^+$, $A^1\Pi$ and $C^1\Pi$ potential energy 
  curves (solid and dashed black curves) in the Hund's case (a) 
  representation that are coupled by the   
  spin-orbit interaction and the resulting  
  $\Omega=1$ relativistic states (red dotted curves) 
  in the Hund's case (c) representation 
  of the SrYb dimer.}   
\label{fig:curves} 
\end{figure} 

The $a^3\Pi$, $b^3\Sigma^+$, $A^1\Pi$, and $C^1\Pi$ excited states essential for the  
photoassociative formation of the ground state SrYb molecule proposed in the next  
section are plotted in Fig. \ref{fig:curves}. The matrix elements of the spin-orbit  
coupling were calculated for the manifolds of coupled $a^3\Pi$, $b^3\Sigma^+$, $A^1\Pi$  
states, cf. Fig. \ref{fig:SOcoup}. The knowledge of the spin-orbit coupling between  
$a^3\Pi$, $b^3\Sigma^+$, $A^1\Pi$, and $C^1\Pi$ states allows us to obtain the relativistic  
$(1)0^-$, $(2)0^-$, $(1)0^+$, $(1)1$, $(2)1$,$(3)1$, $(4)1$ and $(1)2$ states by  
diagonalizing the appropriate relativistic Hamiltonian matrices. The $\Omega=1$ states  
are also plotted in Fig. \ref{fig:curves}. Note that the crossing of the $b^3\Sigma^+$ and  
$A^1\Pi$ nonrelativistic states becomes an avoided crossing between the (2)1 and (3)1 states. 
 
Having all the results briefly presented above we are ready to discuss the photoassociation 
process of cold Sr and Yb atoms, and look for the prospects of producing ultracold SrYb 
molecules. To conclude this section we would like to emphasize that almost all {\em ab initio} 
results were obtained with the most advanced size-consistent methods of quantum 
chemistry: CCSD(T) and LRCCSD. In all calculations all electrons, except for those 
described by the pseudopotentials, were correlated (42 for ytterbium and 10 for 
strontium). Only the SO coupling matrix elements 
and the nonadiabatic matrix elements were obtained with the MRCI method which is 
not size consistent. Fortunately enough, all of the couplings are important 
in the region of the curve crossings or at large 
distances, so the effect of the size-inconsistency of MRCI on our results 
should not be dramatic.

\section{\label{sec:3}Photoassociation and formation of ground state 
  molecules}   
 
Photoassociation is considered for a continuous-wave  
laser that is red-detuned with 
respect to the intercombination line. This transition is 
dipole-forbidden. However, the $a^3\Pi$ state correlating to the 
asymptote of the intercombination line transition, 
cf. Fig.~\ref{fig:curves},  
is coupled by the spin-orbit interaction to 
two singlet states, $A^1\Pi$ and $C^1\Pi$,  
that are connected by a dipole-allowed transition to the 
ground electronic state, $X^1\Sigma^+$. Thus an effective 
transition matrix element is created which can be written, to a very  
good approximation, as  
\begin{eqnarray} 
  \label{eq:dlr} 
  d_{SO}(\Op R) &&=  
\frac{\langle X^1\Sigma^+|\Op{d}|C^1\Pi\rangle\langle C^1\Pi|\Op H_{SO}|a^3\Pi\rangle}
{E_{a^3\Pi}-E_{C^1\Pi}}\nonumber \\
&&+
\frac{\langle X^1\Sigma^+|\Op{d}|A^1\Pi\rangle\langle A^1\Pi|\Op H_{SO}|a^3\Pi\rangle}
{E_{a^3\Pi}-E_{A^1\Pi}}\,,
\end{eqnarray} 
where $\Op H_{SO}$ is the spin-orbit Hamiltonian in the Breit-Pauli 
approximation \cite{Bethe57}. The long-range part of $d_{SO}(\Op R)$, 
dominated by the first term in the above expression, is  
due to the coupling with the $C^1\Pi$ state, ideally 
suited for photoassociation. The short-range part is due to  
the coupling with the $A^1\Pi$ state, paving the way toward efficient 
stabilization of the photoassociated molecules to the electronic 
ground state. The $a^3\Pi$ state, 
in addition to the spin-orbit coupling with the two 
singlet states, is also coupled to the 
$b^3\Sigma^+$ state  correlating to the same asymptote, 
Sr$(^3P)$+Yb$(^1S)$. The Hamiltonian describing these couplings  
yielding the Hund's case (c) $\Omega=1$ states reads 
in the rotating-wave approximation  
\begin{equation}\label{eq:H} 
 \Op H= 
 \begin{pmatrix} 
 \Op H^{X^1\Sigma^+} &  0 & 0  & \half d_{1}(\Op R) E_0 & \half d_{2}(\Op R)E_0  \\ 
 0    & \Op{H}^{a^3\Pi} & \xi_1(\Op R)              & \xi_2(\Op R)           & \xi_4 (\Op R)\\ 
 0    & \xi_1(\Op R)      & \Op{H}^{b^3\Sigma^+}& \xi_3  (\Op R)         & \xi_5 (\Op R)\\ 
\half d_1(\Op R) E_0  & \xi_2  (\Op R)       & \xi_3    (\Op R)          & \Op{H}^{A^1\Pi}& 0 \\ 
\half d_2(\Op R)E_0  & \xi_4   (\Op R)  & \xi_5    (\Op R)   & 0               & \Op{H}^{C^1\Pi}     
 \end{pmatrix}\,, 
\end{equation} 
where $\Op{H}^{^{2S+1}|\Lambda|}$ is the Hamiltonian for  
nuclear motion in the ${^{2S+1}|\Lambda|}$ electronic state, 
$\Op{H}^{^{2S+1}|\Lambda|}=\Op{T}+V^{^{2S+1}|\Lambda|}(\Op R) + V_{\rm trap}^{^{2S+1}|\Lambda|}(\Op R) - (1-\delta_{n0})\hbar\omega_L$. The  
kinetic energy operator is given 
by $\Op{T}={\Op P^2}/{2\mu}$ with $\mu$ the reduced mass of SrYb.  
The trapping potential, $V_{\rm trap}^{^{2S+1}|\Lambda|}(\Op R)$, is relevant only in the 
electronic ground state for the detunings considered below, even for 
large trapping frequencies. We 
approximate it by a harmonic potential which is well justified for 
atoms cooled down to the lowest trap states. The 
parameters of the photoassociation laser are the frequency, $\omega_L$, 
and the maximum field amplitude, $E_0$. 
The electric transition dipole moments are 
denoted by $d_1(\Op R)=\langle X^1\Sigma^+|\Op d|A^1\Pi \rangle$, 
$d_2(\Op R)=\langle X^1\Sigma^+|\Op d|C^1\Pi \rangle$, and the matrix 
elements of the spin-orbit coupling are given by  
\begin{eqnarray*} 
\xi_1(\Op R) &=& \langle a^3\Pi(\Sigma=0,\Lambda=\pm 1) |\Op H_{SO}|b^3\Sigma^+(\Sigma=\pm 1,\Lambda=0)\rangle\,, \\ 
\xi_2(\Op R) &=& \langle a^3\Pi(\Sigma=0,\Lambda=\pm 1) |\Op H_{SO}|A^1\Pi(\Sigma=0,\Lambda=\pm 1) \rangle\,, \\ 
\xi_3(\Op R)&=&\langle b^3\Sigma^+(\Sigma=\pm 1,\Lambda=0) |\Op H_{SO}|A^1\Pi(\Sigma=0,\Lambda=\pm 1) \rangle\,,\\ 
\xi_4(\Op R)&=&\langle a^3\Pi(\Sigma=0,\Lambda=\pm 1) |\Op H_{SO}|C^1\Pi(\Sigma=0,\Lambda=\pm 1) \rangle\,,\\ 
\xi_5(\Op R)&=&\langle b^3\Sigma^+(\Sigma=\pm 1,\Lambda=0) |\Op H_{SO}|C^1\Pi(\Sigma=0,\Lambda=\pm 1) \rangle\,. 
\end{eqnarray*}   
$\Sigma$ and $\Lambda$ denote the quantum numbers for the projections of 
electronic spin and orbital angular momenta,  $\Op{S}$ and  $\Op{L}$, 
onto the internuclear axis.  
Note that the specific shape of the $C^1\Pi$ potential energy curve as 
well as the $R$-dependence of its spin-orbit coupling and transition 
dipole matrix elements are not important, since the $C^1\Pi$ state 
provides the effective dipole coupling only at long range. We have 
therefore approximated the $R$-dependence of the couplings with the 
$C^1\Pi$ state by their  
constant asymptotic values in the calculations presented below. 
The Hamiltonian \eqref{eq:H} has been represented on a Fourier grid 
with adaptive step size~\cite{SlavaJCP99,WillnerJCP04,ShimshonCPL06}  
(using $N=1685$  grid points and grid mapping parameters 
$\beta=0.22$, $E_{\rm min}=7\cdot 10^{-9}\,$hartree).  
 
\begin{figure}[tb] 
\centering 
\includegraphics[width=0.95\linewidth]{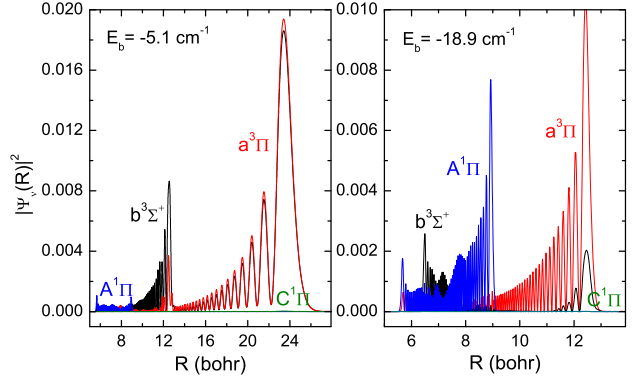} 
\caption{Vibrational wave functions of the coupled $a^3\Pi$, 
  $b^3\Sigma^+$, $A^1\Pi$, and $C^1\Pi$ electronic states of SrYb 
  molecule for two binding energies corresponding to vibrational 
  quantum numbers $v'=-11$ (left) and $v'=-18$ (right) 
  below the dissociation  threshold.  
}  
\label{fig:Psi} 
\end{figure} 
\begin{figure}[tb] 
\centering 
\includegraphics[width=1\linewidth]{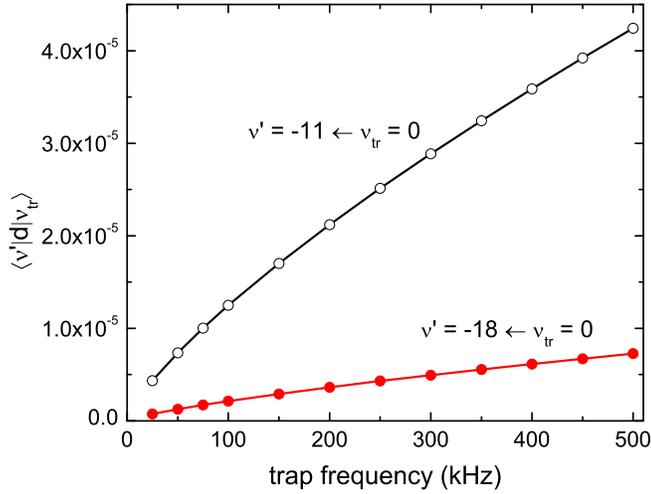} 
\caption{Vibrationally averaged free-to-bound (or quasi-bound-to-bound) 
  electric transition dipole moments 
  between the lowest trap state of a pair of Sr and Yb atoms  
  colliding in the $X^1\Sigma^+$ ground 
  electronic state in a  harmonic trap and two bound levels, 
  cf. Figure~\ref{fig:Psi}, of  electronically excited  
  SrYb dimers as a function of the trap frequency. 
}  
\label{fig:FCFHz} 
\end{figure} 
The key idea of photoassociation using a continuous-wave laser 
is to excite a colliding pair of 
atoms into a bound level of an electronically excited 
state.~\cite{FrancoiseReview,JonesRMP06} 
For maximum photoassociation efficiency,  
the detuning of the laser with respect to the atomic asymptote, 
Sr$(^3P_1)$+Yb$(^1S)$ in our case, is chosen to coincide with the 
binding energy of one of the vibrational levels in the electronically 
excited state. Fig.~\ref{fig:Psi} shows two such levels with binding 
energies $E_b=5.1\,$cm$^{-1}$ (left) and $E_b=18.9\,$cm$^{-1}$ (right). 
Since four electronically excited states are coupled by the spin-orbit 
interaction, the vibrational wavefunctions have components on all four 
electronically excited states. Note that the norm of the 
$C^1\Pi$-component of these two vibrational wavefunctions is smaller 
than $10^{-3}$.  
Nevertheless, this is enough, similar to the photoassociation of the strontium 
dimers near an intercombination line,~\cite{ZelevinskyPRL06} 
to provide the transition dipole for the free-to-bound (or 
quasi-bound-to-bound, due to the trapping potential) excitation.  
The vibrational level with binding energy $E_b=5.1\,$cm$^{-1}$ is 
predominantly of triplet character (with 56\% of its norm residing on 
the $a^3\Pi$ state, 32\% on the $b^3\Sigma^+$ state and just 11\% 
on the $A^1\Pi$ state), while  
the vibrational level with binding energy $E_b=18.9\,$cm$^{-1}$ shows a truly 
mixed character (55\% triplet vs 45\% singlet).  
The fact that multiple classical turning points are clearly visible in  
the vibrational wavefunction with $E_b=18.9\,$cm$^{-1}$ reflects the resonant 
nature of the spin-orbit coupling of this 
level: the coinciding energy of the levels in the coupled vibrational 
ladders leads to a resonant beating between the different components 
of the coupled wavefunctions.~\cite{AmiotPRL99}  
Such a structure of the vibrational 
wavefunctions was shown to be ideally suited for efficient 
stabilization of the photoassociated molecules into deeply bound levels 
in the ground electronic   
state.~\cite{DionPRL01,PechkisPRA07,Ghosal09,LondonoPRA09}  
 
The Condon radius for photoassociation 
coincides with the classical outer turning point, i.e.,  
roughly speaking with the outermost peak of the vibrational 
wavefunctions as shown in Fig.~\ref{fig:Psi}.  
Since the pair density of the atoms colliding in their electronic 
ground state  
decreases with decreasing interatomic distance, photoassociation is 
more efficient for small detuning. 
This is reflected by the larger values of the black compared to the 
red curve in Fig.~\ref{fig:FCFHz} which shows the free-to-bound 
(quasi-bound-to-bound)  
transition matrix elements for the two vibrational wavefunctions 
depicted in Fig.~\ref{fig:Psi} as a function of the trapping frequency 
of the optical lattice. The second observation to be drawn from 
Fig.~\ref{fig:FCFHz} is the almost linear scaling of the transition 
matrix elements, and hence the photoassociation probability, with the 
trap frequency. That is, enhancing the trap frequency from 
50$\,$kHz, which has been employed for photoassociation of 
Sr$_2$~\cite{ZelevinskyPRL06},  to 
500$\,$kHz, which is within current experimental feasibility, 
will increase the number of photoassociated molecules by a half an  
order of magnitude.  
This confinement effect is easily understood in 
terms of the larger compression of the quasi-bound atom pairs in a 
tighter optical trap.  
 
\begin{figure}[tb] 
\centering 
\includegraphics[width=0.95\linewidth]{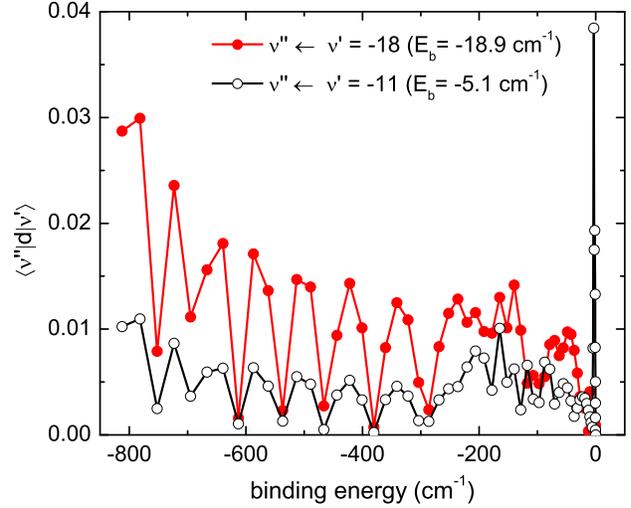} 
\caption{Vibrationally averaged bound-to-bound  
  electric transition dipole moments 
  between the vibrational levels of the coupled  
  electronically excited states that are shown in Fig.~\ref{fig:Psi}  
  and all vibrational levels of the $X^1\Sigma^+$ ground  
  electronic state.}  
\label{fig:FCF} 
\end{figure} 
In view of the formation of deeply bound molecules in their electronic 
ground state, it might be advantageous to choose the larger detuning  
of $18.9\,$cm$^{-1}$ 
despite the photoassociation probability being smaller by about a factor of 
5.9 compared to a detuning of $5.1\,$cm$^{-1}$ 
for all trap frequencies. 
This becomes evident by inspecting Fig.~\ref{fig:FCF} which displays the 
bound-to-bound transition matrix elements between the two electronically 
excited vibrational wavefunctions with $E_b=5.1\,$cm$^{-1}$ and 
$E_b=18.9\,$cm$^{-1}$ and all bound levels of the $X^1\Sigma^+$ 
electronic ground state. These transition matrix elements govern the 
branching ratios for spontaneous decay of the photoassociated 
molecules. Note that for $v^\prime=-11$ and   
$v^\prime=-18$, the electronically excited molecules will 
decay into bound levels of the electronic ground state with a 
probability of about 24\%. 
This decay to a large extent into bound levels  
is a hallmark of photoassociation near an intercombination 
line.~\cite{ZelevinskyPRL06} It is in contrast to photoassociation 
using a dipole-allowed transition where the probability for 
dissociative decay is often several orders of magnitude 
larger than that for stabilization into 
bound ground state levels.~\cite{FrancoiseReview} 
While the excited state vibrational level with $E_b=5.1\,$cm$^{-1}$ 
has its largest transition dipole matrix elements with the last 
bound levels of the $X^1\Sigma^+$ ground electronic state that are 
only weakly bound, a striking 
difference is observed for the excited state vibrational wavefunction 
with $E_b=18.9\,$cm$^{-1}$.  
The strong singlet-triplet mixing of this level, in particular the 
pronounced peak near the  
outer classical turning point of the $A^1\Pi$ state, 
cf. Fig.~\ref{fig:Psi}, leads to significantly stronger transition 
dipole matrix elements with deeply bound levels of the  
$X^1\Sigma^+$ ground electronic state for $v^{\prime}=-18$ compared to 
$v^{\prime}=-11$, the one with $v^{\prime\prime}=1$  
being the largest. Of course, the transition dipole matrix elements 
govern not only the spontaneous decay of the photoassociated molecules 
but also stabilization via stimulated emission. Due to the 
comparatively long lifetime of photoassociated molecules, estimated to 
be of the order of $15\,\mu$s, stabilization into a selected single 
vibrational level of the electronic ground state can be achieved 
by stimulated emission using a second continuous-wave laser.  
 
\begin{figure}[tb] 
\centering 
\includegraphics[width=0.95\linewidth]{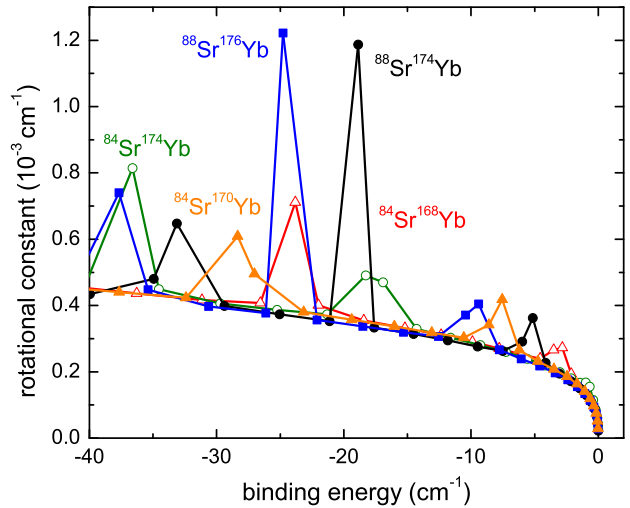} 
\caption{Rotational constants of the vibrational levels 
  of the coupled $a^3\Pi$, $b^3\Sigma^+$, $A^1\Pi$, and $C^1\Pi$ 
  electronically excited states of the SrYb molecule for different 
  isotope pairs. The isotope $^{88}$Sr$^{174}$Yb was employed in the 
  calculations shown in Figs.~\ref{fig:Psi}-\ref{fig:FCF} and \ref{fig:Scheme}. }   
\label{fig:Rotational} 
\end{figure}

Before outlining how a prospective experiment forming SrYb molecules 
in their vibronic ground state based on our results could proceed,  
it is natural to ask whether the accuracy of the calculations is 
sufficient for such a prediction. In particular,  
how sensitively do our results for the binding 
energies and structure of the vibrational levels as well as for the 
transition matrix elements depend on the accuracy of the electronic 
structure calculations?  
The binding energies depend mostly on the quality of the potential 
energy curves, where the error is estimated to be a few percent, and 
to some extent, for the spin-orbit coupled excited states, on the 
accuracy of the spin-orbit interaction matrix elements (error of 
a few percent). 
Nevertheless, the uncertainty of our potential energy curves is 
smaller than the range of reduced masses, as illustrated in 
Fig.~\ref{fig:Rotational}.  Therefore 
photoassociation with subsequent 
stabilization to a low-lying vibrational level should work for all 
isotope pairs since levels with strong perturbations due to 
the spin-orbit interaction are always present  
in the relevant range of binding energies, 
respectively, detunings, cf. Fig.~\ref{fig:Rotational}. 
 
In fact, the exact position and the character of the excited state 
vibrational level,  
strongly perturbed such as the one with $E_b=18.9\,$cm$^{-1}$ or 
more regular such as that with $E_b=5.1\,$cm$^{-1}$ in 
Fig.~\ref{fig:Psi}, can be determined 
experimentally.~\cite{PechkisPRA07,FiorettiJPB07}   
A possible spectroscopic signature of the character of the vibrational  
wavefunctions is the dependence of the rotational constants, $\langle 
v^\prime|\frac{1}{2\mu\Op R^2}|v^\prime\rangle$, on the 
binding energy of the corresponding levels. This is shown in 
Fig.~\ref{fig:Rotational} for  
different isotope combinations of strontium and ytterbium. The 
rotational constants of those levels that are predominantly of triplet 
character lie on a smooth curve, while those that are mixed deviate 
from this curve. This behavior is easily understood as follows: 
Without the coupling due to the spin-orbit interaction, the rotational 
constants of the $a^3\Pi$, $b^3\Sigma^+$ and $A^1\Pi$ states would each 
lie on a smooth curve. The strongly mixed levels 'belong' to all three  
curves at the same time. Correspondingly, the value of their 
rotational constant lies somewhere inbetween the smooth curves of the 
regular levels. The lower peaks at small binding energies in 
Fig.~\ref{fig:Rotational} indicate mixing mostly between the  
$a^3\Pi$ and $b^3\Sigma^+$ states, while the high peaks at larger 
binding energies reflect a strong singlet-triplet mixing.  
Spectroscopic determination of the rotational constants thus allows for 
identifying those excited state levels that show the strongest 
singlet-triplet mixing and are best suited to the 
formation of ground state molecules.  
Spectroscopy is also needed to refine the value for the transition 
frequency of the stabilization laser. 
The binding energies of vibrational levels of the $X^1\Sigma^+$ 
ground electronic state come with an error of  
5~\%, i.e., $\pm$~50~cm$^{-1}$, defining the window for 
spectroscopic search. 
 
Note that our model, Eq.~\eqref{eq:H}, does not account for angular  
couplings, i.e., couplings of the $\Omega=1$ states with  
$\Omega=0^{\pm}$ and $\Omega=2$. 
When including non-adiabatic angular couplings, we found 
the components of  
the vibrational wavefunctions on the newly coupled surfaces  
to account for less than 0.001\% of the 
population. The changes in the binding energy of the vibrational 
levels turned out to be less than $10^{-6}$cm$^{-1}$, well within the error 
of the electronic structure calculations. This negligible effect of 
the angular (Coriolis-type) couplings for SrYb is not surprising  
due to its large reduced mass which enters inversely all coupling 
matrix elements. 
 
\begin{figure}[tb] 
\centering 
\includegraphics[width=0.95\linewidth]{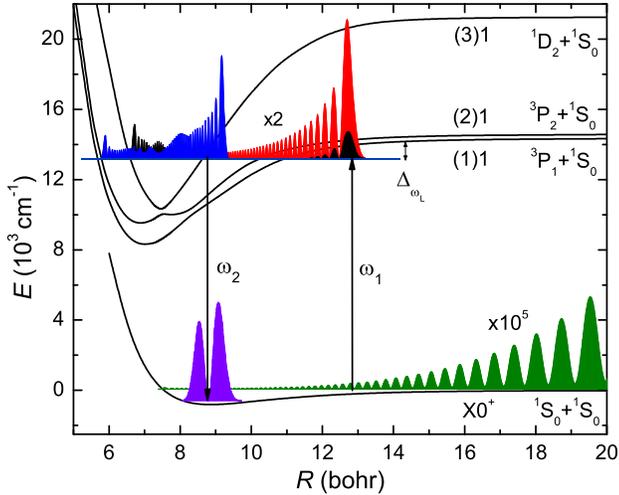} 
\caption{Proposed scheme for the formation of ground state SrYb 
  molecules via photoassociation near the intercombination line 
  transition with detuning $\Delta_{\omega_L}=18.9\,$cm$^{-1}$  
  ($\nu_\mathsf{trap}=100\,$MHz).} 
\label{fig:Scheme} 
\end{figure} 
Combining all results shown above and assuming that the relevant 
spectroscopic data has been confirmed or adjusted experimentally,  
we suggest the following scheme for 
photoassociation of SrYb dimers followed by stabilization via 
stimulated emission (see Fig.~\ref{fig:Scheme}):  
\begin{enumerate} 
\item A large trapping 
frequency of the optical lattice is chosen to optimally compress the 
pair density of strontium and ytterbium atoms prior to 
photoassociation.  
\item A photoassociation laser with frequency $\omega_1\approx 
  690\,$nm, red-detuned from the  
  intercombination line transition and resonant with an electronically 
excited vibrational level, $v^\prime$, of strongly mixed singlet-tripled 
character, is applied for a few $\mu$s. The duration of the 
photoassociation laser (about $5\,\mu$s roughly is an upper bound) 
is a compromise between saturating photoassociation and avoiding 
spontaneous emission losses (lifetime of about 15$\,\mu$s) while the 
laser is on. 
\item As the photoassociation laser is switched off, the stabilization laser 
  is switched on. Due to the strong bound-to-bound transition matrix 
  elements, saturation of the transition is expected already for shorter 
  pulses ($\le 1\,\mu$s). The frequency of the stabilization laser, 
  $\omega_2\approx 655\,$nm, is chosen to be resonant with the 
  transition from the electronically excited level, $v^\prime$, to the 
  first excited vibrational level of the $X^1\Sigma^+$ electronic ground 
  state, $v^{\prime\prime}=1$.  
\item Before repeating steps 2 and 3, both photoassociation and 
  stabilization lasers remain turned off for a hold period in which 
  the $X^1\Sigma^+(v^{\prime\prime}=1)$ molecules decay to the vibronic 
  ground state, $X^1\Sigma^+(v^{\prime\prime}=0)$. This ensures that 
  the molecules created in the electronic ground state by the first 
  sequence of the   
  photoassociation and stabilization steps are not re-excited in a 
  following sequence. The formed molecules can then be accumulated in  
  $X^1\Sigma^+(v^{\prime\prime}=0)$.  
\end{enumerate} 
Note that this scheme does not require phase coherence between the two 
pulses.  
Step 4 needs to involve a dissipative element in order to ensure the 
unidirectionality of the molecule formation scheme.~\cite{KochMoszPRA08} 
Dissipation can be 
provided by infrared spontaneous emission due to the permanent dipole 
moment of the heteronuclear dimers. However, this timescale is 
estimated to be of the order of 5$\,$s, much too slow to be 
efficient for accumulation of ground state molecules. A second 
possibility is due to collisional decay. For the decay to occur within 
$1\,$ms, a density of $10^{13}\,$cm$^{-3}$ is required.  
Note that the density was  $3\cdot 10^{12}\,$cm$^{-3}$ in the 
experiment photoassociating Sr$_2$ in an optical lattice with trapping 
frequency 50$\,$kHz.~\cite{ZelevinskyPRL06} Increasing the trap 
frequency will further increase the density such that hold times in 
the sub-ms regime are within experimental reach. 
 
One might wonder whether the comparatively long hold times can be 
avoided by using Stimulated Raman Adiabatic Passage 
(STIRAP)~\cite{BergmannRMP98} for the photoassociation (pump) and 
stabilization (Stokes) pulses.~\cite{ShapiroPRA07,KuznetsovaPRA08}  
In order to overcome the problem of unidirectionality that occurs in 
repeating the photoassociation and stabilization steps many times,  
the whole ensemble of atom pairs in the trap needs to be addressed  
within a single STIRAP sweep~\cite{KuznetsovaPRA08}  
or within a single sequence of phase-locked STIRAP 
pulse pairs.~\cite{ShapiroPRA07}  
Note that the Stokes/stabilization pulse should be tuned to the 
$v^\prime\to v^{\prime\prime}=0$ transition in this case. 
The feasibility of STIRAP-formation of ground state molecules  depends 
on isolating the initial state sufficiently from  
the scattering continuum. A possibility to achieve this  
that was discussed theoretically consists in 
utilizing the presence of a Feshbach 
resonance.~\cite{KuznetsovaPRA08,KuznetsovaNJP09} If no resonance is 
present, i.e., in an unstructured scattering continuum, STIRAP fails.  
In a series of ground-breaking experiments, STIRAP transfer to the 
ground state was therefore preceded by Feshbach-associating the  
molecules.~\cite{LangPRL08,OspelkausNatPhys08,NiScience08,DanzlSci08}    
An alternative way to isolate the  initial state for STIRAP from  
the scattering continuum that does not rely on Feshbach resonances 
(which are absent for the even isotope species of Sr and Yb) 
is given by strong confinement in a deep optical lattice. An estimate 
of the required trap frequency is given in terms of  
the binding energy of the Feshbach molecules that were 
STIRAP-transferred to the vibronic ground state. It was for example 
about $230\,$kHz for KRb molecules.~\cite{OspelkausNatPhys08,NiScience08}  
Hence a deep optical lattice with trapping frequency of the order of a 
few hundred kHz should be sufficient to enable  
STIRAP-formation of ground state molecules. However, in order to be 
adiabatic with respect to the vibrational motion in the trap with 
periods of the order of about $1\,\mu$s, the duration of the 
photoassociation pulse needs to rather long, at least of the order of 
$10\,\mu$s. The challenge might be to maintain phase coherence 
between the photoassociation pulse and the stabilization pulse over 
such timescales. For a train of phase-locked STIRAP-pulse 
pairs,~\cite{ShapiroPRA07}  the requirement of durations of the order 
of $10\,\mu$s or larger applies to the length of the sequence of pulse 
pairs. The minimum Rabi frequencies to enforce adiabatic following are 
$\Omega =159\,$kHz for a $10\,\mu$s-pulse or $\Omega=15.9\,$kHz for 
a $100\,\mu$s-pulse. 
As a further prerequisite, all 
or at least most atom pairs should reside in the lowest trap state, 
$v_{\mathsf{trap}}=0$. Then 
steps 2-4 above might be replaced, provided the trapping 
frequency is sufficiently large, by 
\begin{enumerate} 
\item[2.$^\prime$] a single STIRAP-sweep~\cite{BergmannRMP98}  
  forming ground state molecules with $\mu$s-pulses where the 
  stabilization laser, tuned on resonance with the $v^\prime\to 
  v^{\prime\prime}=0$ transition ($\omega_2\approx 654\,$nm),  
  precedes the photoassociation laser, 
  tuned on resonance with the $v_{\mathsf{trap}}=0 \to 
  v^{\prime}$  transition ($\omega_2\approx 690\,$nm);   
\item[2.$^{\prime\prime}$] or,  
  a train of short, phase-locked  
  STIRAP pulse pairs with correctly adjusted pulse 
  amplitudes.~\cite{ShapiroPRA07} 
\end{enumerate} 
To convert the Rabi frequencies to field amplitudes, note that the 
transition matrix elements are $5\cdot 10^{-6}$ for the pump pulse 
(assuming a trap frequency of 300$\,$kHz) and $3\cdot 10^{-2}$ for the 
Stokes pulse. Phase coherence needs to be maintained 
throughout the single STIRAP-sweep or sequence of STIRAP pulse pairs. 
 
\section{\label{sec:4}Summary} 
 
Based on a first principles study,  
we predict the photoassociative formation of SrYb molecules in their 
electronic ground state using transitions near an intercombination 
line.  
The potential energy curves, non-adiabatic angular 
coupling and spin orbit interaction matrix elements as well as 
electric dipole transition matrix elements of the SrYb molecule were 
calculated with 
state-of-the-art \textit{ab initio} methods, using the coupled cluster 
and multireference configuration interaction frameworks.  
Assuming that the accuracy of the calculations for the SrYb molecule 
is about the same as for the isolated Sr and Yb atoms at the same 
level of the theory, we  estimate 
the accuracy of the electronic structure data to 5\%. 
However, the crucial point for the proposed photoassociation scheme  
is the existence and position of the 
intersection of the potential energy curves corresponding to 
$b^3\Sigma$ and $A^1\Pi$ states. By contrast to the binding 
energies of the vibrational levels, the position of this intersection 
does not depend very much on the overall quality of the computed potential 
energy curves curves. The correct structure of the crossings between 
the potential curves of the $a^3\Pi$, $b^3\Sigma$ and $A^1\Pi$ states  
is reproduced using even relatively crude computational methods of 
quantum chemistry which  
do not account for dynamic correlations such as the multiconfiguration 
self-consistent field (MCSCF) method employed here.

The spin-orbit coupled $a^3\Pi$, $b^3\Sigma^+$, $A^1\Pi$, and $C^1\Pi$ 
electronically excited states are essential for the 
photoassociation. A pair of colliding Sr and Yb atoms is excited into 
the triplet states with the dipole coupling for the photoassociation 
(stabilization) provided by the  $C^1\Pi$ ($A^1\Pi$) state. 
The formation of SrYb molecules in their electronic ground state  
proceeds via photoassociation into a weakly bound level of the coupled 
electronically excited states ($\omega_1\approx 690\,$nm) 
followed either by spontaneous or 
stimulated emission.  
 
If photoassociation is followed by spontaneous emission,  
about 24\% of the photoassociated molecules will 
decay into bound levels of the ground electronic state, roughly 
independent of the detuning of the photoassociation  
laser. However, \textit{which} ground state rovibrational levels are populated by spontaneous 
emission depends strongly on the detuning of the photoassociation 
laser. While most detunings will lead to decay into the last bound 
levels of the ground electronic states, certain detunings populate  
excited state levels with strong spin-orbit mixing. The strongly  
resonant structure of the wavefunctions allows for decay into 
low-lying vibrational levels. This might be the starting point for 
vibrational cooling~\cite{AllonCP01,PilletSci08}  
if molecules in their vibronic ground state are 
desired.  
 
Alternatively, the long lifetime of the photoassociated 
molecules, of the order of $15\,\mu$s, allows for stabilization to the 
electronic ground state via stimulated emission, 
by a sequence of photoassociation and stabilization laser pulses of 
$\mu$s duration. Two schemes are conceivable: (i) A repeated cycle of  
photoassociation and stabilization pulses is applied with 
$X^1\Sigma^+(v^{\prime\prime}=1)$ as the target 
level. The duration of the pulses should be of the order of $1\,\mu$s.   
In order to  
accumulate molecules in $X^1\Sigma^+(v=0)$, a hold period whose 
duration depends on the density of atoms is 
required for collisional decay from $v=1$ to $v=0$.  
For deep optical lattices, hold periods in the sub-ms regime can be 
reached.  
(ii) The vibronic ground state, $X^1\Sigma^+(v=0)$, is 
targeted directly by a counter-intuitive sequence of photoassociation 
and stabilization pulses (STIRAP),  
either using two long pulses~\cite{BergmannRMP98} 
or a train of phase-locked pulse pairs.~\cite{ShapiroPRA07}  
The timescale for the pulses is determined by the requirement to be 
adiabatic with respect to the motion in the optical lattice. The 
largest trapping frequencies feasible to date  
imply pulse durations at least of the order of 
$10\,\mu$s. Phase coherence between the pulses needs to be maintained 
over this timescale.  
Note that STIRAP fails if applied to an unstructured scattering 
continuum of colliding atoms. A possibility to circumvent this is 
given by  
preselecting the initial state for STIRAP with the help of a 
(Feshbach) resonance.~\cite{ShapiroPRA07,KuznetsovaPRA08,KuznetsovaNJP09}   
Our variant of the scheme is different since  
STIRAP is enabled by the presence of a deep trap.    
 
Before either of the above discussed  molecule formation 
schemes can be implemented experimentally, our 
theoretical data needs to be corroborated by spectroscopy. In 
particular, our binding energies come with an error of a few percent, 
implying a corresponding uncertainty in the transition 
frequencies. Moreover, the exact position of strongly spin-orbit mixed 
excited state wavefunctions needs to be confirmed by measuring 
the excited state level spacings or rotational constants. 
However, despite the relatively large uncertainties in the energies of the  
rovibrational levels important for the proposed photoassociation scheme,  
our {\em ab initio} methods correctly locate the crossing  
of the singlet and triplet potential energy curves. This is the key 
ingredient for the efficient production of ground state SrYb 
molecules that we are predicting with our study. 
 
\section*{Acknowledgments} 
 
We would like to thank Tatiana Korona and Wojciech Skomorowski for 
many useful discussions and help with the \textsc{Molpro} program.  
This work was supported by the Polish Ministry of Science and 
Education through the project N N204 215539, and by the  
Deutsche Forschungsgemeinschaft (Grant No. KO 2301/2). 
MT was supported by the project operated within the Foundation for 
Polish Science MPD Programme co-financed by the EU European Regional 
Development Fund.

 
 
\providecommand*{\mcitethebibliography}{\thebibliography}
\csname @ifundefined\endcsname{endmcitethebibliography}
{\let\endmcitethebibliography\endthebibliography}{}

\end{document}